\documentclass[twocolumn,showpacs,superscriptaddress,floatfix,prl]{revtex4-1}
\usepackage{graphicx,amsfonts,amssymb,amsmath,hyperref,hypcap,enumerate}

\usepackage{amsthm}
\usepackage{multirow}
\usepackage{graphicx}
\usepackage{dcolumn}   
\usepackage{bm}        
\usepackage{amssymb}   
\usepackage{amsmath}

\DeclareMathAlphabet{\mathitb}{OT1}{cmr}{bx}{sl}

\begin{document}
\title{ Chiral Topological Excitons in a Chern Band Insulator  }

\author{Ke Chen}
\affiliation{International Center for Quantum Materials, Peking University, Beijing 100871, China}
\affiliation{Collaborative Innovation Center of Quantum Matter, Beijing 100871, China}

\author{Ryuichi Shindou}
\email{rshindou@pku.edu.cn}
\affiliation{International Center for Quantum Materials, Peking University, Beijing 100871, China}
\affiliation{Collaborative Innovation Center of Quantum Matter, Beijing 100871, China}

\date{\today}

\begin{abstract}
A family of semiconductors called as Chern band insulator are shown to 
host exciton bands with non-zero topological Chern integers and chiral exciton 
edge modes. 
Using a prototypical two-band Chern insulator model, 
we calculate a cross-correlation function 
to obtain the exciton bands and their Chern integers.
The lowest exciton band acquires Chern
integers such as $\pm 1$ and $\pm 2$ in electronic Chern insulator phase. 
The non-trivial topology can be experimentally 
observed both by non-local optoelectronic response of
exciton edge modes and by a phase shift in the cross-correlation response due to the 
bulk mode. Our result suggests that magnetically 
doped HgTe, InAs/GaSb quantum wells and $\text{(Bi,Sb)}_{2} \text{Te}_{3}$ thin film are 
promising candidates for a platform of topological excitonics. 

\end{abstract}


\maketitle

Exciton is an electron-hole bound state in semiconductors, which plays central roles  
in semiconductor optoelectronics. A binding energy of the exciton becomes dramatically  
enhanced in low-dimensional semiconductors due to quantum confinement effect~\cite{efros1,brus,keldysh}.  
Well-studied examples are 
excitons in quantum dot~\cite{ekimov,efros2,norris}, wire~\cite{arakawa,asada},   
carbon nanotube~\cite{heinz,maul,dress} and two-dimensional 
materials such as transition metal dichalcogenide (TMDC) 
monolayer~\cite{feng,chei,qiu,komsa,shi,ram,xu,zeng,lagarde,mai,wang,yu1,cao,jones,pospischil,baugher,ross}.  

\begin{figure}[t]
        \includegraphics[width=1\columnwidth]{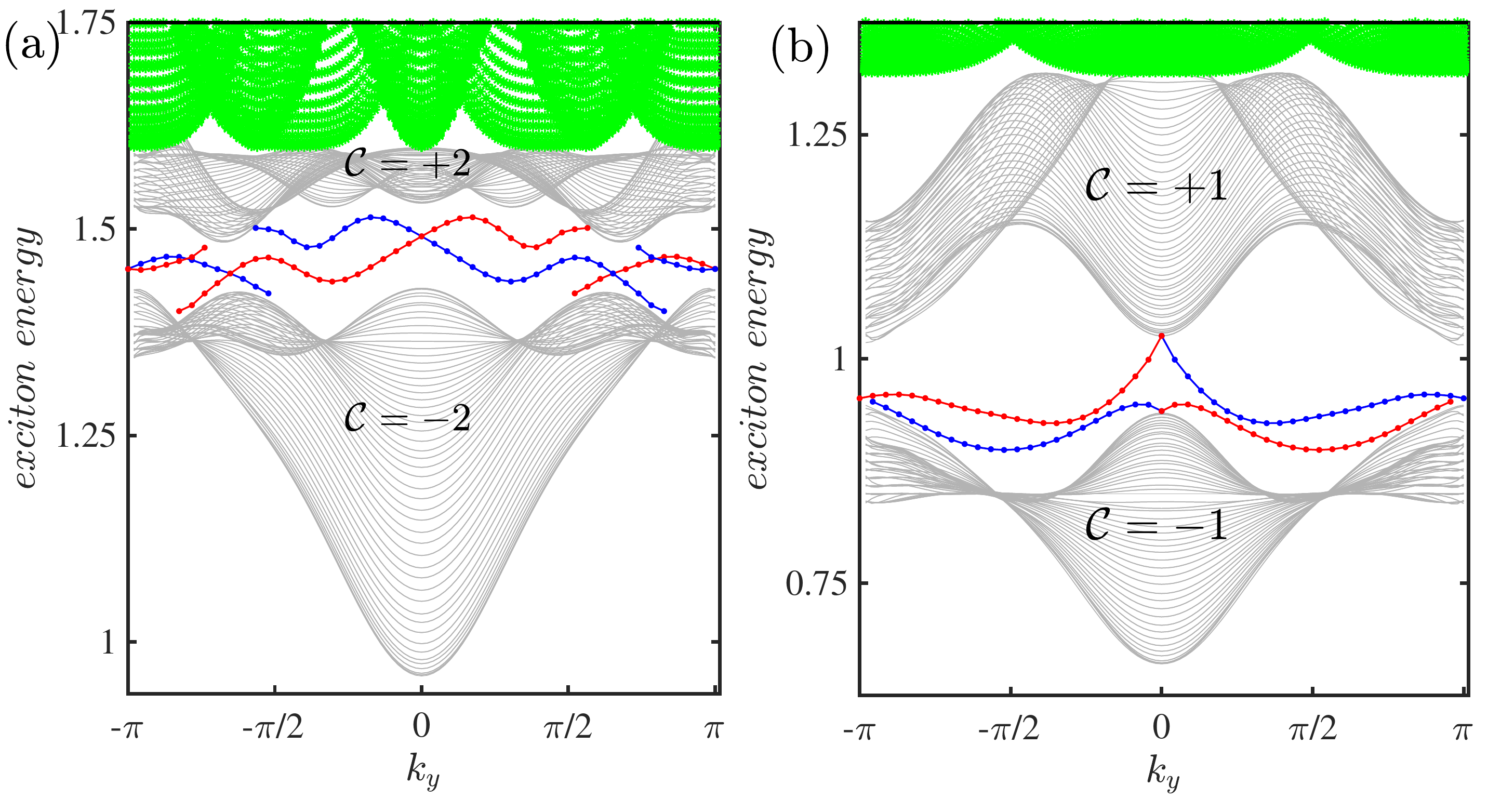}
        \caption{(color online) Chiral exciton edge modes and exciton bulk modes (open/periodic 
boundary condition in $x$/$y$-axis). The blue and red line represent edge modes 
localized at opposite boundaries ($x=0$ and $x=30$ respectively). The green-color 
region is an electron-hole continuum due to transitions between bulk valence 
and conduction bands.   
        (a) $U=3.31,t'=0.5,t=1.5,m=1$ (``$c$''-point in Fig.~\ref{Fig:pd}).    
        (b) $U=3.31,t'=0.5,t=1.16,m=0.8$ (``$d$''-point in Fig.~\ref{Fig:pd}). 
        }
        \label{Fig:edge}
        
\end{figure}

Topological excitonics in low-dimensional semiconductors
~\cite{jain,seradjeh1,seradjeh2,budich,yuenzhou1,yuenzhou2,karzig,nalitov,bardyn,wangyao,yu2,wu1,wu2,gong} 
offers unique perspective in optoelectronics and future energy-harvesting materials. 
A topological exciton edge mode has an energy-momentum dispersion 
within the light cone and interacts with light (Fig.~\ref{Fig:edge}).
Under {\it p}-{\it n} junction, such topological exciton edge modes enable  
strong electroluminescence (EL) with much longer exciton life time: the 
strong EL intensity is due to spatially localized nature of the edge exciton wavefunction 
and the longer life time can be associated with its limited decay process due to 
a peculiar topological protection of the modes. Unidirectional nature of topological 
chiral exciton edge modes enables novel {\it non-local} 
optoelectronic response. 
The first theoretical proposal was made 
in organic semiconductors~\cite{yuenzhou1,yuenzhou2}.  
Thereby, dipolar interactions among Frenkel excitons play vital roles in realization of 
topological excitonics like in topological magnonics~\cite{shindou1}.   
Synthetic gauge field in photon-exciton couplings 
endows polaritons with non-trivial band topology~\cite{karzig,nalitov,bardyn}. 
A bright exciton with a Dirac cone spectrum in TMDC 
monolayer~\cite{yu1,wangyao,yu2,wu1,jhzhou} is 
theoretically suggested to realize topological exciton  
under Moir\'e patterns or moderate strains with external magnetic field~\cite{wu2,gong}.   

Chern insulator is a two-dimensional topological band insulator with broken time reversal symmetry  
in which the quantized Hall conductance is realized without external magnetic field~\cite{haldane1}. The first 
material realization was proposed theoretically in magnetic atoms doped two-dimensional quantum 
spin Hall insulators~\cite{cxliu1,cxliu2,cxliu3}.  Later, an experimental realization 
was achieved in a thin film of magnetic topological insulator, 
Cr-doped (Bi,Sb)$_{2}$Te$_{3}$~\cite{chang}, where a magnetic field dependence of the 
two-dimensional Hall conductance clearly shows the quantized Hall conductance 
of $\pm e^2/h$ in the zero external field. 
 
 In this paper, we demonstrate that the Chern band insulator provides unprecedented opportunity 
to explore rich topological exciton physics. We show that a prototypical model for the Chern band insulator 
hosts exciton bands with non-zero topological integers and topological chiral exciton edge modes that are 
bright. We study a two-band square lattice model for the Chern band insulator with an inclusion of on-site 
Coulomb repulsion term, to calculate a linear response function among density and 
pseudospin degree of freedoms. Eigenvalues of the matrix-formed response  
function have a well-defined pole below electron-hole continuum, which 
describes an energy-momentum dispersion for exciton excitations in the Chern insulator. 
We define topological Chern integer for the exciton bands from the 
corresponding eigenvectors. We reveal that the lowest and second lowest 
exciton bands acquire a variety of non-zero Chern integers  
within a parameter region for the Chern insulator phase. Consequently, the integer number
of chiral exciton modes localized in boundaries run across 
the band gap between these exciton bulk bands. From their wavefunctions and dispersions, 
they are bright excitons. We argue that the non-trivial band 
topology of the exciton bulk bands can be directly 
mapped out by a measurement of a phase shift 
of the cross-correlation encoded in the matrix-formed response function.

{\em The model.}--- The first material realization of Chern band insulator 
is proposed in Mn atoms doped two-dimensional quantum spin Hall systems 
such as HgTe and InAs/GaSb quantum wells~\cite{cxliu1,cxliu3,cxliu2}. 
When ferromagnetically ordered (magnetic moment upward), the doping magnetic atoms induce 
exchange fields both in conduction electron band ($s$-wave band) and hole band 
($p$-wave band) but in an opposite direction with each other, which renders a band inversion 
between the electron band with down spin ($\downarrow$)  and the hole band with up spin ($\uparrow$) 
to be reinverted, while leaving intact the band inversion between the other pair. 
This leads to a low-energy effective two-band model for the Chern insulator~\cite{qi}. In the momentum 
space, ${\bm k}\equiv (k_x,k_y)$, the kinetic energy part of the Hamiltonian on square lattice   
takes a form of  ${\cal H}_0 \equiv \sum_{\bm k} {\bm c}^{\dagger}_{\bm k} \!\ {\bm H}_{sp}({\bm k}) 
\!\ {\bm c}_{\bm k}$ 
with 
\begin{align}
 {\bm H}_{sp} ({\bm k})  & \equiv  -t' \left( \cos k_{x}  + \cos k_{y}  \right) {\bm \sigma}_{0} 
+ \sin k_{x} \!\ {\bm \sigma}_{1}   \nonumber \\ 
& +  \sin k_{y} \!\ {\bm \sigma}_{2} 
+ \left( m - t \left(\cos  k_{x} + \cos k_{y} \right) \right) \!\ {\bm \sigma}_{3} \label{hami}
\end{align}
and ${\bm c}^{\dagger}_{\bm k} \equiv (c^{\dagger}_{{\bm k},s,\uparrow},c^{\dagger}_{{\bm k},p,\downarrow})$. 
${\bm \sigma}_a$ is a $ 2\times2 $ Pauli matrix composed by $s$ orbital with $\uparrow$ and 
$p_{+} \equiv p_x+ip_y$ orbital with $\downarrow$.  $-t-t^{\prime} \!\ (<0)$ and  
$t-t^{\prime} \!\ (>0)$ are the nearest neighbor intra-orbital hopping integrals 
for $s$ and $p_{+}$ orbitals respectively. $m$ is an atomic energy difference between 
the two orbitals. For $|m|<|2t|$, an inter-orbital hopping due to the relativistic spin-orbit 
interaction induces a band gap, making the system to 
be QAH insulator (Chern insulator)~\cite{qi}. 
The inter-orbital hopping with odd spatial parity dictates that uniform electric currents 
$J_x$ and $J_y$ contain pseudospin density components: 
$ J_{\mu}=\sum_{\bm{k}}{\bm{c}}_{\bm{k}}^{\dagger}\cos k_{\mu}{\bm{\sigma}}_{\mu}{\bm{c}}_{\bm{k}}+\cdots $ 
 ($ \mu=x(1),y(2) $).  
We take the inter-orbital hopping to be unit. As for 
a screened Coulomb interaction, we consider an on-site Coulomb interaction $U \!\ (>0)$ 
for simplicity:
\begin{align}
{\cal V} &=\frac{1}{2}\frac{U}{N^{2}}\sum_{{\bm k}_{1},{\bm k}_{2},{\bm q},\alpha,\beta} 
c_{{\bm k}_{1}+{\bm q},\alpha}^{\dagger}c_{{\bm k}_{2}-{\bm q},\beta}^{\dagger}c_{{\bm k}_{2},\beta}c_{{\bm k}_{1},\alpha}
\label{hami2}
\end{align}   
with $\alpha,\beta=1,2$ which stand for  
$(s,\uparrow)$ and $(p,\downarrow)$ respectively. $N^2$ is total number 
of the square lattice sites. 
 
\begin{figure}[t]
	\includegraphics[width=0.9\columnwidth]{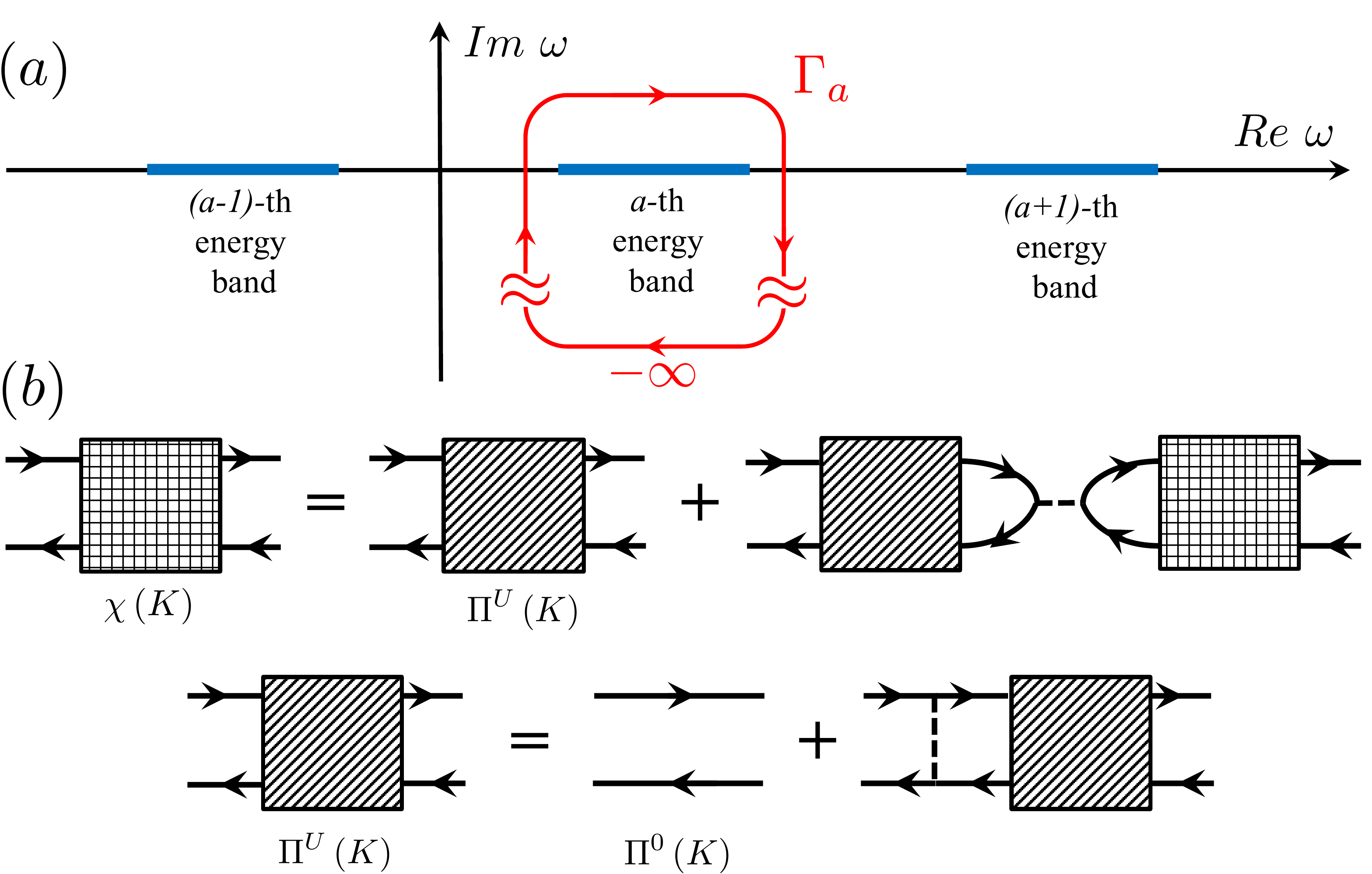}
	\caption{(color online) 
	(a) Closed loop along which an integral over complex $\omega$ is taken in Eq.~(\ref{Ca}).
	(b) Feynman diagrams for the correlation function within 
the generalized random phase approximation. ${\bm \Pi}^{U}$: two-particle irreducible Green's 
function, ${\bm \Pi}^{0}$: bare polarization part.}
	\label{Fig:fd}
\end{figure} 

{\em Response function and Topological integer.}--- 
The Chern integer for exciton bulk band is a central building block of topological 
excitonics. We define this in connection with linear response of the system against 
external perturbations. Consider an external field ${\bm J}_b({\bm r}_m)$ which couples with the density ($a=0$)
and pseudospin degree of freedom ($a=1,2,3$): ${\cal O}_{a}({\bm r}_j) \equiv c^{\dagger}_{{\bm r}_j,\alpha} 
({\bm \sigma}_{a})_{\alpha\beta} c_{{\bm r}_j,\beta}$. The density and pseudospin densities induced 
by the external fields are given by a linear response function as 
$\langle {\cal O}_a({\bm r}_j,t) \rangle = \int d{\bm r}_m \!\ 
\int^{\infty}_{-\infty} dt^{\prime} \chi^{R}_{ab}({\bm r}_j-{\bm r}_m,t-t^{\prime}) 
{\bm J}_{b}({\bm r}_m,t^{\prime})$ with $\chi^{R}_{ab}({\bm r}_j-{\bm r}_m,t-t^{\prime}) 
\equiv -i\theta(t-t^{\prime}) \langle [{\cal O}_{a,H}({\bm r}_j,t), 
{\cal O}_{b,H}({\bm r}_m,t^{\prime})]\rangle$. The response function in the dual space, 
$ \ensuremath{\chi_{ab}^{R}({\bm{k}},\omega)\equiv\int dtd{\bm{r}}\!\ 
e^{-i{\bm{k}}\cdot{\bm{r}}+i\omega t}\chi_{ab}^{R}({\bm{r}},t)} $, 
is directly related to the optical conductivity in some cases. For example,  
when bright excitons are composed mainly by particle-hole pairs near the time-reversal 
symmetric momentum points, $\ensuremath{\chi_{11}^{R}} $, 
$\ensuremath{\chi_{12}^{R}} $ and $ \ensuremath{\chi_{22}^{R}}$ associated with 
these excitons (divided by frequency $ \ensuremath{\omega}$) contribute directly 
to optical conductivities, $\ensuremath{\sigma_{xx}} $, $\ensuremath{\sigma_{xy}}$ 
and $\ensuremath{\sigma_{yy}} $ respectively. 
  
From an analogy of the quantum Hall physics~\cite{tknn,kohmoto,ishikawa1,ishikawa2},  
the Chern integer for the $a$-th bulk exciton band (`$a$' is an index for the bulk band) 
is defined by the response function in the dual space:
\begin{align}
C_a \equiv \frac{\epsilon_{\mu\nu}}{2} \int_{\rm BZ} \frac{{\bm d}{\bm k}}{2\pi} 
\oint_{\Gamma_a} \frac{d\omega}{2\pi} {\rm Tr}\bigg[\frac{\partial {\bm \chi}^R}{\partial \omega} 
\frac{\partial\left({\bm{\chi}}^{R}\right)^{-1}}{\partial k_{\mu}} 
{\bm \chi}^R \frac{\partial\left({\bm{\chi}}^{R}\right)^{-1}}{\partial k_{\nu}} \bigg],  
\label{Ca}
\end{align}
Here an integral over $\omega$ is along a loop $\Gamma_a$ which encompasses an energy 
region of the $a$-th bulk band on the real $\omega$ axis (Fig.~\ref{Fig:fd}). 
In the dilute electron and hole density limit~\cite{fetter,hanke,maiti}, the response 
function can be calculated by a generalized random phase approximation 
(Fig.~\ref{Fig:fd}),
\begin{align}
\chi^{R}_{ab}({\bm k},\omega) &= \chi^{T}_{ab}({\bm k},i\omega_n=\omega+i\eta), \label{r-t} \\ 
\chi^T_{ab}(K) & =  \Pi_{ab}^{U}(K)+\frac{U \Pi_{a0}^{U}(K) \!\ \Pi_{0b}^{U}(K)}{1-U\Pi_{00}^{U}(K)},  
\label{rpa} \\ 
{\bm \Pi}^{U}(K) & =  {\bm \Pi}^{0}(K)  \!\ 
\Big[{\bm 1}_{4\times4}+\frac{U}{2}{\bm \Pi}^{0}(K)\Big]^{-1},  \label{ladder} \\ 
\Pi_{ab}^{0}(K) &\equiv \frac{1}{\beta}\frac{1}{N^{2}}\sum_{Q}{\rm Tr}\left[{\bm \sigma}_{a} {\bm g}^{0} (Q+K) 
{\bm \sigma}_{b} {\bm g}^{0}(Q)\right], \label{bare} 
\end{align} 
with $K\equiv({\bm k},i\omega_{n})$ and $Q\equiv({\bm q},i\epsilon_{n})$. A $ 2\times2 $  
bare electron Green's function ${\bm g}^{0}(Q)$ is given by eigenvectors of 
${\bm H}_{sp}({\bm q})$, $|{\bm q},j\rangle$: 
\begin{align}
{\bm g}^{0}\left({\bm q},i\epsilon_{n}\right) &\equiv 
\sum_{j=c,v}\frac{|{\bm q},j\rangle \langle {\bm q},j|}{i\epsilon_{n}-{\cal E}_{{\bm q},j} +\mu} \nonumber 
\end{align}
with ${\bm H}_{sp}({\bm q}) |{\bm q},j\rangle = {\cal E}_{{\bm q},j} |{\bm q},j\rangle$. 
$j=c,v$ denotes conduction and valence band respectively. 
Recast in the picture of effective single-exciton Hamiltonian~\cite{pikus,maialle}, the 
first term in Eq.~(\ref{rpa}) (ladder diagrams) corresponds to the direct interaction between electron and hole, while 
the second term corresponds to the exchange interaction. 
For simplicity of the calculation, we have considered only the ladder diagrams. 
 
Below the electron-hole continuum, $\omega<{\rm min}_{\bm q}({\cal E}_{{\bm q}+{\bm k},c}-{\cal E}_{{\bm q},v})$, 
the response function ${\bm \chi}^{R}({\bm k},\omega)$ is Hermitian and is 
diagonalized by a unitary matrix, unless one of its eigenvalues has a pole. Namely, 
\begin{align}
\left[{\bm{\chi}}^{R}({\bm{k}},\omega)\right]^{-1}|u_{a}({\bm{k}},\omega)\rangle
=|u_{a}({\bm{k}},\omega)\rangle L_{a}({\bm{k}},\omega), \nonumber 
\end{align}
where $|u_{a}({\bm k},\omega)\rangle$ with $a=0,1,2,3$ form an orthonormal basis for 
those $\omega$ with $L_a({\bm k},\omega) \ne 0$ for all $a$. The eigenvalue has at 
most one pole for each $a$ in the energy region below the electron-hole continuum,  
$\omega=E_{a,{\bm k}}$ with $L_a({\bm k},E_{a,{\bm k}})=0$, which determines an energy-momentum 
dispersion of the $a$-th exciton bulk band. Near each pole, 
the response function takes an asymptotic form 
\begin{align}
{\bm \chi}^{R}({\bm k},\omega) = |\tilde{u}_a({\bm k})\rangle \frac{A_{a}({\bm k})}{\omega-E_{a,{\bm k}}+i\eta} 
\langle \tilde{u}_{a}({\bm k})| + \cdots. \label{chiR}
\end{align} 
$|\tilde{u}_{a}({\bm k})\rangle \equiv |u_a({\bm k},E_{a,{\bm k}})\rangle$ is a 
Bloch wavefunction for the $a$-th bulk exciton band and 
$A_{a}^{-1}({\bm{k}})\equiv\frac{\partial L_{a}}{\partial\omega}|_{\omega=E_{a,{\bm{k}}}}$ 
is the inverse of a 
spectral weight. Eqs.~(\ref{Ca},\ref{chiR}) give the Chern integer in terms of the 
Berry curvature defined by the Bloch wavefunction $|\tilde{u}_a\rangle$: 
$C_a \equiv \frac{1}{\pi}\int d{\bm k} \!\ {\rm Im} \big( \partial_{k_y} \langle \tilde{u}_a |\big) 
\big( \partial_{k_x} |\tilde{u}_a \rangle\big)$~\cite{haldane2,shindou-4}.  

\begin{figure}[t]
	\includegraphics[width=0.9\columnwidth]{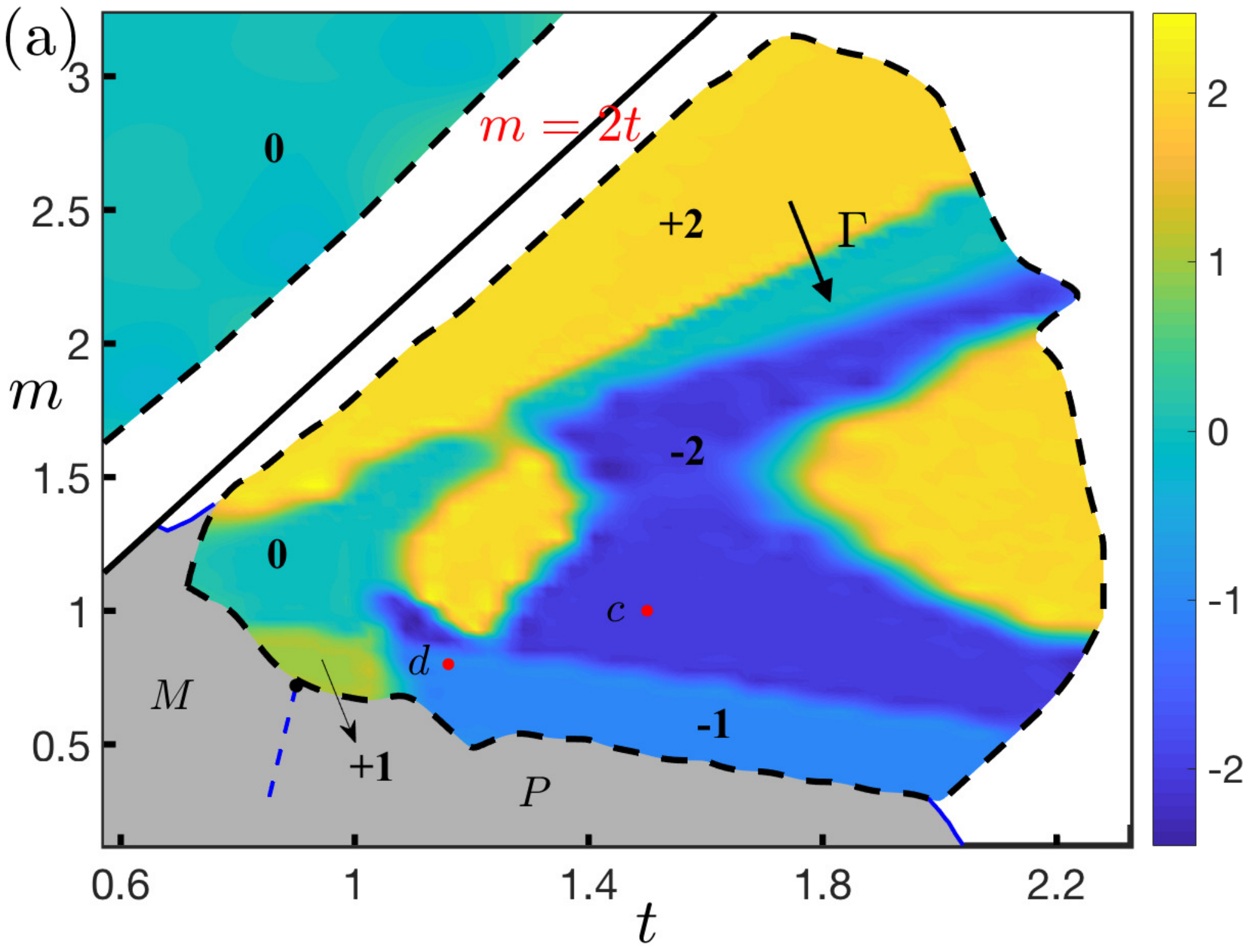}
	\includegraphics[width=0.9\columnwidth]{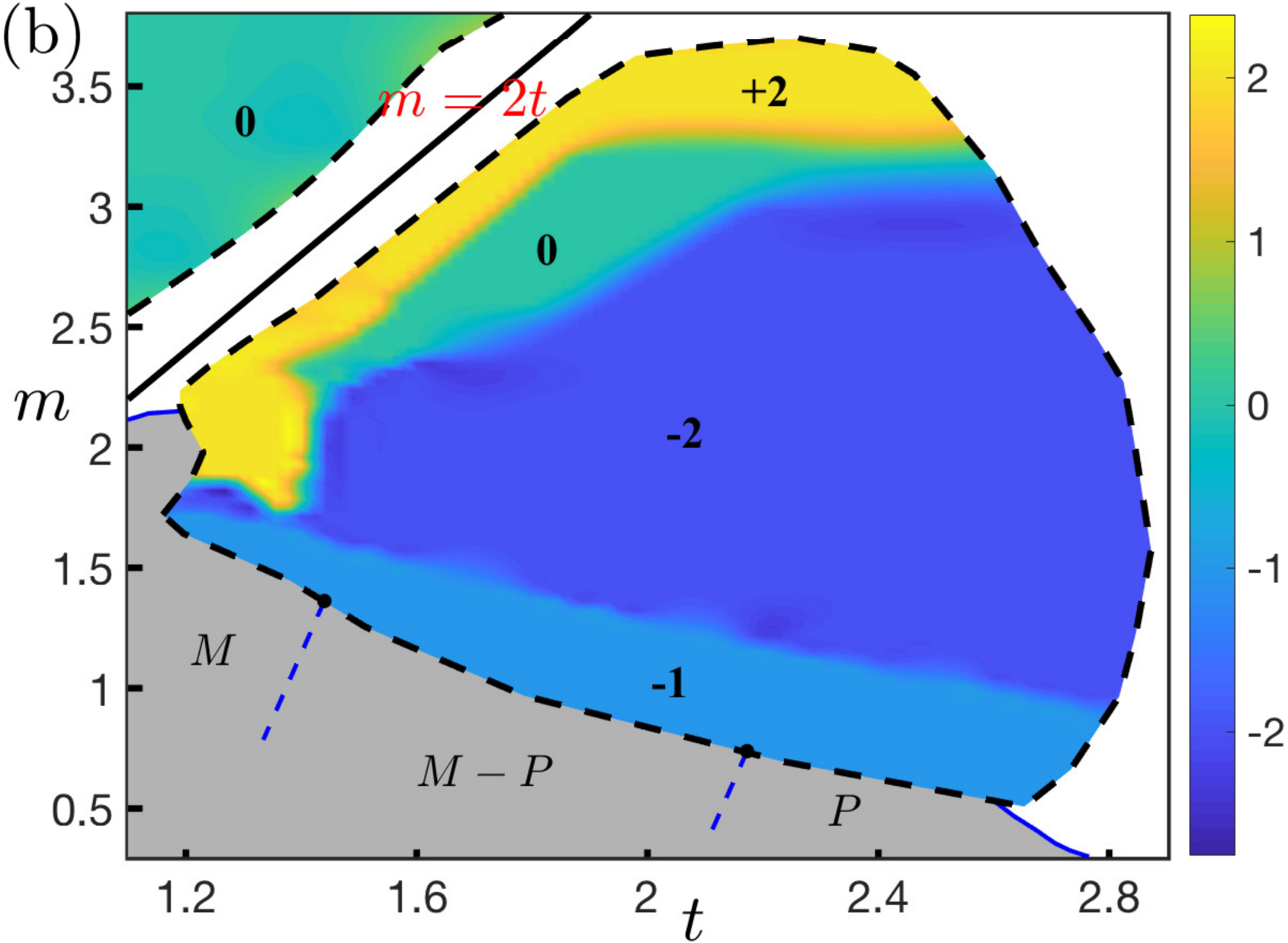}
	\caption{(color online) Distributions of the Chern integer of the lowest exciton bulk 
band in $(t,m)$ parameter space at $U=3.31, t'=0.5$ (a) and at $U=3.95$, $t'=1.0$ (b). Regions with 
colors (yellow, green, blue, yellow green, light blue) and with Chern integers 
($+2,0,-2,+1,-1$) represent those parameter regions where the lowest exciton bulk band is well 
separated from the electron-hole continuum in the whole Brillouin zone (BZ). In the white region, 
the band enters the continuum at certain points of the BZ. When the parameters are tuned from 
the colored regions to the grey region, the lowest exciton band touches the zero energy, giving 
rise to an exciton condensation. The labels `$M$', `$P$' and `$ M-P $' indicate that
the exciton condensation occours at the $M$, $P$, and points between $M$ and $P$ point respectively.}
	\label{Fig:pd}
\end{figure}

{\em Phase diagram.}---
For moderately large $U$ ($2 \lesssim U \lesssim 5$), the lowest and second lowest 
exciton bands are well separated from the electron-hole continuum. Such exciton 
bands almost always take a variety of non-zero Chern integers in the Chern 
insulator phase region ($|m|<2|t|$). Fig.~\ref{Fig:pd} shows a distribution of the Chern integer of the lowest exciton band 
around $U=3$ and $4$. When the integer changes from one to another, 
the two exciton bands form a linear or quadratic band touching. The transitions from $+2$ to 
$0$ (labelled in Fig.~\ref{Fig:pd}(a)), from $0$ to $-2$, and from $-2$ to $-1$ region are 
accompanied by a quadratic band touching at $\Gamma$ point, linear touching at 
two $M$ points, ${\bm k}=(0,\pi)$, $(\pi,0)$, and linear touching 
at $P$ point, ${\bm k}=(\pi,\pi)$, respectively. 

Effective $2\times2$ Hamiltonians for these exciton band touchings can be 
derived from a symmetry argument of a Bethe-Salpeter (BS) Hamiltonian for single 
exciton wavefunction~\cite{sup}. 
The BS Hamiltonian is symmetric under a magnetic point
group $4m^{\prime}m^{\prime}$. 
The eigen wavefunctions of the BS Hamiltonian 
at $\Gamma$ and $P$ point form an irreducible corepresentation (IcREP) of 
$4m^{\prime}m^{\prime}$. There are four distinct 1-dimensional IcREPs of 
$4m^{\prime}m^{\prime}$
under which the eigen wavefunction  
has $s$, $d$, $p_x+ip_y$ and $p_x-ip_y$-wave symmetry respectively. 
The exciton band touchings are composed by two states belonging to distinct IcREPs. 
An analysis shows that the quadratic band touching at $\Gamma$ point is composed by 
two odd-parity states ($p_{+}$ and $p_{-}$) or two even-parity states ($s$ and $d$)~\cite{sup}.
The linear band touching at $P$ point 
is by even and odd states (e.g. $s$ and $p_{+}$). 

For the ordinary insulator region ($|m|>2|t|$), we observe only single exciton bulk 
band with zero Chern integer below the electron-hole continuum: being 
consistent with an ionic limit ($m \gg 2t$), where the BS Hamiltonian is completely 
real-valued.  
When $U$ gets much larger than the band gap (e.g. $U\gtrsim 5$), 
the lowest exciton bulk band reaches the zero energy at high symmetric points 
such as $M$ and $P$ points, only to exhibit an exciton condensation (Fig.~\ref{Fig:pd}). 
The resulting electronic phase for $|m|<2|t|$ is another band insulator
with reduced translational symmetry.   
 
{\em Nature of topological chiral exciton edge modes.}--- 
In the region with non-zero Chern integers in Fig.~\ref{Fig:pd}, the 
response function is calculated for the system 
with open boundary condition in one direction ($x$) and periodic 
condition in the other ($y$): $\chi^{R}_{ab}(x_j,x_m;k_y,\omega)$ with 
$k_y$ momentum conjugate to $y$~\cite{sup}. When an indirect gap opens 
between the lowest two exciton bulk bands, the response function acquires several 
new poles inside the gap. 
The poles give chiral dispersions as a function of 
$k_y$, connecting the two exciton bulk bands (Fig.~\ref{Fig:edge}). 
We observe that the number of chiral dispersions with 
sign ($+$ for left and $-$ for right-handed) is consistently 
identical to Chern integer of the lowest exciton band ${\mathbb Z}$~\cite{halperin,hatsugai}.   
The eigen wavefunctions of the response function which 
correspond to these new poles are all localized at the spatial boundaries. They have significant 
weights in the pseudospin components of ${\bm \sigma}_1$ and 
${\bm \sigma}_2$, representing chiral exciton edge modes that interact with light. 

\begin{figure}[t]
	\includegraphics[width=0.95\columnwidth]{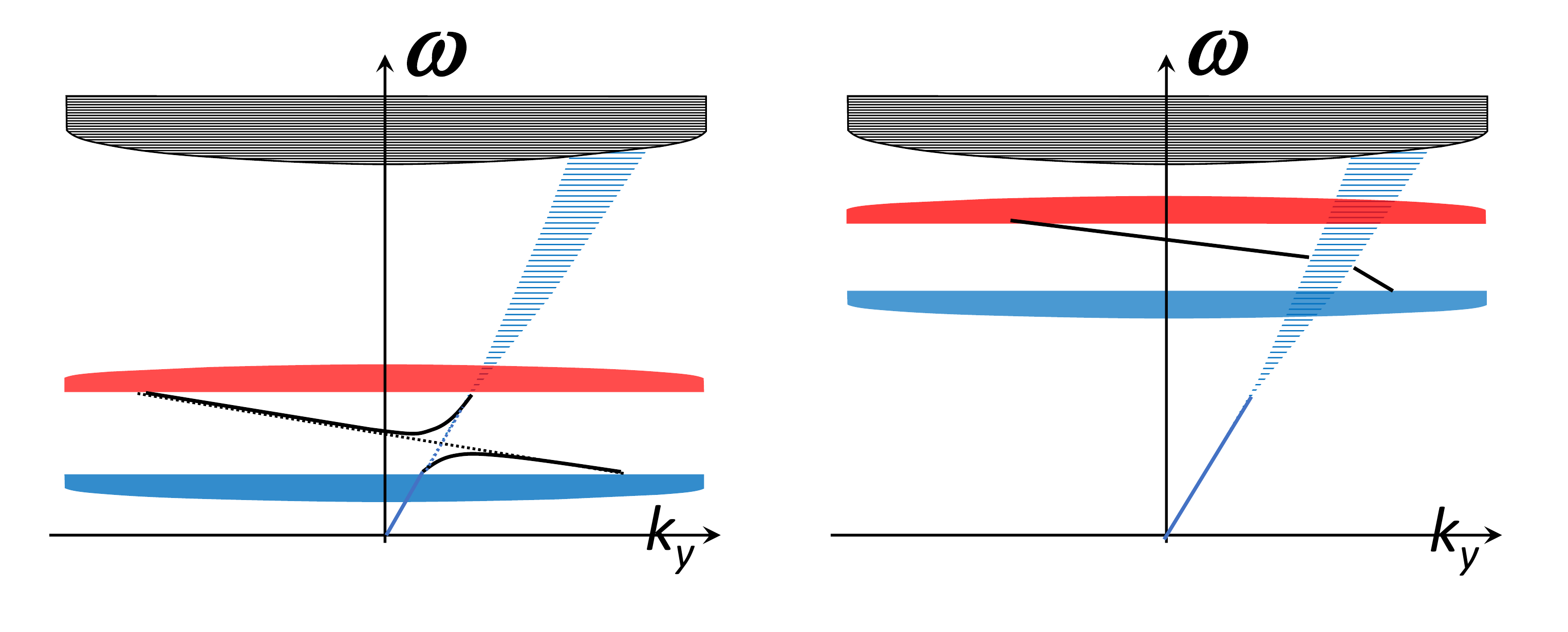}
	\caption{(color online) Schematic pictures of level repulsion between chiral phason mode 
and exciton edge mode (left), and chiral exciton edge mode decaying in edge electron-hole continuum (right). 
Electron-hole excitations along the gapless chiral electronic edge mode form the chiral 
phason mode (blue solid line) in low-energy region and edge electron-hole continuum (blue shaded region) 
in high-energy region. Black solid line, black shaded region, blue region and 
red region stand for chiral exciton edge mode, bulk electron-hole continuum, 
the lowest and 2nd lowest exciton bulk bands respectively.}
	\label{Fig:ee}
\end{figure}

The Chern band insulator has a gapless electronic edge mode  
with right and left-handed chiral dispersion for $0<m<2t$ 
and $-2t<m<0$ respectively~\cite{qi}. In low-energy region, 
electron-hole excitations along the electronic edge mode 
form an 1-dimensional collective excitation 
(chiral phason mode)~\cite{giamarchi}, which can have level crossings with low-energy exciton edge 
modes. Being not protected by symmetry, 
the crossings lead to level repulsions (Fig.~\ref{Fig:ee}). Thus, 
total number of chiral dispersions of the {\it low-energy} edge excitons 
inside the indirect gap becomes ${\mathbb Z}-1$ for $0<m<2t$ and 
${\mathbb Z}+1$ for $-2t<m<0$ instead of ${\mathbb Z}$ 
(${\mathbb Z}$ 
also changes sign under $m\rightarrow -m$). 
In high-energy region (lower than the electron-hole continuum 
associated with the bulk states), the electron-hole excitations along the 
edge provide another continuum due to finite 
curvature of the electronic edge mode dispersion. 
This edge electron-hole continuum gives a   
finite life time to the chiral exciton edge modes, 
when they share energy and momentum (Fig.~\ref{Fig:ee}).  

{\em Phase shift in the cross response function.}---
When the exciton bulk band has a non-zero 
Chern integer, $C_a \ne 0$, there always exists a momentum ${\bm k}_{0,j}$ in the BZ 
at which $\langle j|\tilde{u}_{a}\rangle=0$ and around which the component 
has a U(1) phase winding 
$\langle j|\tilde{u}_{a}\rangle \propto e^{i\theta_{{\bm k},j}}$ with 
$\oint_{|{\bm k}-{\bm k}_{0,j}|=\epsilon} 
{\bm \nabla}_{\bm k} \theta_{{\bm k},j} \cdot d{\bm k} = {\mathbb Z}$ for 
any $j=0,1,2,3$~\cite{tknn,kohmoto}. 
The phase winding can be directly 
mapped out by a phase-sensitive spectroscopic measurement of ${\bm \chi}^R$,  
e.g., a phase-sensitive Brillouin light scattering~\cite{serga,schneider,demokritov}. 
Thereby, an incident probe light wave 
excites exciton at $a$-th bulk band with momentum ${\bm k}$, 
 which leads to an absorption at $\omega=E_{a,{\bm k}}$.  
 Eq.~(\ref{chiR}) dictates that 
the absorption spectrum in the cross-correlation component 
$ \chi_{jl}^{R}\left(\bm{k},\omega=E_{a,\bm{k}}\right) $ ($ \ensuremath{j\ne l} $) 
acquires a phase shift of $\Delta \theta_{\bm k} \equiv \theta_{{\bm k},j}-\theta_{{\bm k},l}$ 
with $\langle l |\tilde{u}_a\rangle \propto e^{i\theta_{{\bm k},l}}$. 
 When ${\bm k}$ goes around the boundary of an area $S$ in the 
${\bm k}$ space that encloses ${\bm k}_{0,j}$, 
the phase shift shows a U(1) phase holonomy of $2\pi {\mathbb Z}$. 
The phase 
holonomy is determined only by the number of the vortex of $ \theta_{{\bm k},j} $
within the area $S$ minus that of $ \theta_{{\bm k},l} $.  

{\em Conclusion.}--- 
A prototypical Chern insulator model with an on-site Coulomb interaction  
hosts exciton bands with non-zero topological Chern integers such as $\pm1$ 
and $\pm2$ in the Chern insulator phase, with consequent chiral exciton edge modes. 
The chiral exciton edge mode may have level repulsion with chiral edge phason mode in low
energy region, while short-wavelength exciton edge mode can be damped into 
edge electron-hole excitations only in a small region of its wavelength.  
The non-trivial band topology can also be observed from a phase shift in 
the cross-correlation response by a phase-sensitive spectroscopic 
measurement. The Chern insulator in magnetic topological insulator thin film~\cite{chang,kung} 
can be described by the same low-energy effective continuous model of Eq.~(\ref{hami})~\cite{rui_yu},   
therefore can host qualitatively similar chiral topological excitons as in this paper.  Our results also 
suggest that other quantum anomalous Hall insulators such as ferromagnetic graphene 
under YIG~\cite{jingshi}  are also potential candidates for seeking topological excitons.

The authors thank Junren Shi for helpful discussions.
This work was supported by NBRP of China Grants No.~2014CB920901, 
No.~2015CB921104, and No.~2017A040215. 
 
\bibliographystyle{apsrev4-1}
\bibliography{refs}


\clearpage
\begin{center}
\textbf{Supplemental Material}
\end{center}

\section{Effective Hamiltonians for Exciton Band Touchings}
\subsection{$\Gamma$ and $P$ Points}
Effective Hamiltonians for the exciton band touchings at high symmetric points 
can be derived from a symmetry analysis on an effective equation of motion for 
a single exciton (BS equation)~\cite{s-pikus,s-maialle}.  
We begin with an exciton creation operator:
\begin{align}
b_{a,{\bm k}}^{\dagger} \equiv \sum_{{\bm q}} \phi_{a}\left({\bm q},{\bm k}\right) 
f_{{\bm q}+\frac{\bm k}{2},c}^{\dagger} f_{{\bm q}-\frac{\bm k}{2},v}  
\end{align} 
where $f^{\dagger}_{{\bm q},j}$ denotes a creation operator for conduction band electron ($j=c$)  
or valence band electron ($j=v$). $\phi_a({\bm q},{\bm k})$ is an eigen wavefunction of 
the Bethe-Salpeter equation. For Eqs.~(1,2) in the main text, the eigenvalue problem 
takes a form of 
\begin{align}
& \sum_{{\bm q}'} K({\bm q},{\bm q}';{\bm k}) \!\ \phi_{a}({\bm q}',{\bm k}) 
= E_{a,{\bm k}}\phi_{a}({\bm q},{\bm k}), \\ 
& K({\bm q},{\bm q}^{\prime};{\bm k})
=\delta_{{\bm q},{\bm q}'} ({\cal E}_{{\bm q}+{\bm k}/2,c}-{\cal E}_{{\bm q}-{\bm k}/2,v}) \nonumber \\
& + \frac{U}{N^{2}} \big \langle {\bm q}+\frac{\bm k}{2},c \big|{\bm q}-\frac{\bm k}{2},v 
\big\rangle  \big \langle {\bm q}'-\frac{\bm k}{2},v \big| {\bm q}'+\frac{\bm k}{2},c \big\rangle \nonumber \\
& - \frac{U}{N^{2}}\big\langle {\bm q}+\frac{\bm k}{2},c\big|{\bm q}'+\frac{\bm k}{2},c
\big\rangle \big\langle {\bm q}'-\frac{\bm k}{2},v\big|{\bm q}-\frac{\bm k}{2},v\big\rangle. \label{bs}
\end{align} 
Here $|{\bm q},j\rangle$ and ${\cal E}_{{\bm q},j}$ are an eigenvector and eigenvalue 
of the non-interacting Hamiltonian ${\bm H}_{sp}({\bm q})$.  The Hamiltonian 
is symmetric under a magnetic point group 
$4m^{\prime}m^{\prime}=\{E,C_4,C^2_4,C^{-1}_4,\sigma_{x}T,\sigma_y T, \sigma_{X}T, \sigma_Y T\}$, 
with a 4-fold rotation around $z$-axis $C_4$, a mirror with respect to $\mu z$ 
plane $\sigma_{\mu}$ and time-reversal $T$. Since the interaction part respects the symmetry, 
the BS Hamiltonian $K({\bm q},{\bm q}^{\prime};{\bm k})$ is also symmetric under $4m^{\prime}m^{\prime}$, 
e.g.,  
\begin{align}
&K({\bm q},{\bm q}^{\prime};{\bm k}) = K(C_4({\bm q}),C_4({\bm q}^{\prime});C_4({\bm k})), \nonumber \\ 
&K^{*}({\bm q},{\bm q}^{\prime};{\bm k}) = K(-\sigma_x({\bm q}),-\sigma_x({\bm q}^{\prime});-\sigma_x({\bm k})). \nonumber 
\end{align}   

Eigen wavefunctions of the BS Hamiltonian  
at $\Gamma$ and $P$ points form an irreducible corepresentation 
(IcREP) of $4m^{\prime}m^{\prime}$. There are four distinct 1-dimensional IcREPs of 
$4m^{\prime}m^{\prime}$. They are $A$, $B$, $E_1$, and $E_2$~\cite{s-bradley}, in which the eigen wavefunction 
$\phi_a({\bm q},{\bm k})$ is transformed with $s$, $d$, $p_x+ip_y$ and $p_x-ip_y$-wave 
symmetry under the symmetry operation on ${\bm q}$ (and ${\bm k}$) respectively 
(see Table~I), e.g.,  
\begin{align}
&\phi_{E_1}(C^{-1}_4({\bm q}),{\bm k}=0) = (+i) \phi_{E_1}({\bm q},{\bm k}=0), \nonumber \\ 
&\phi^{*}_{E_1}(-\sigma_X({\bm q}),{\bm k}=0) = (+i) \phi_{E_1}({\bm q},{\bm k}=0). \nonumber 
\end{align}   

\begin{table}[h]
\caption{character table of $4m'm'$}
\begin{center}
\begin{ruledtabular}
\begin{tabular}{lcccccccc}
        & $E$ & $C_{4}^{1}$ & $C_{4}^{2}$ &$C_{4}^{-1}$ & $\sigma_{x}T$ & $\sigma_{y}T$ & $\sigma_{X}T $ & $ \sigma_{Y}T$  \\
\colrule
   $A$ & $1$ &$1$ & $1$& $1$ & $1$ &$1$ &$1$& $1$ \\ 
   $B$ & $1$ &$-1$ & $1$& $-1$ & $1$ &$1$ &$-1$& $-1$ \\ 
   $E_{1}$ & $1$ &$+i$ & $-1$& $-i$ & $1$ &$-1$ &$+i$& $-i$ \\ 
   $E_{2}$ & $1$ &$-i$ & $-1$& $+i$ & $1$ &$-1$ &$-i$& $+i$ \\ 
\end{tabular}
\vspace{0.2cm}

\end{ruledtabular}
\end{center}

\end{table}

The exciton band touchings at the highest symmetric points are composed by those two eigen wavefunctions  
belonging to distinct IcREPs: any level crossings between two eigenstates that belong to a same 
IcREP are generally lifted by symmetry-allowed terms, thus they are not stable band touchings.

A form of the effective $ 2\times2 $ Hamiltonian formed by a pair of two distinct 
IcREPs is determined by the symmetry. As an example, consider that 
$E_1$ ($p_+$) and $E_2$ ($p_-$) states form a band 
touching at $\Gamma$ point and $X_{m}=0$ ($X_{m}$ is a model parameter such as $m$ and $t$ in 
Fig.~3 in the main text). By the ${\bm k}\cdot {\bm p}$ perturbation theory, 
the BS Hamiltonian gives out a $ 2\times2 $ effective Hamiltonian around this degeneracy 
point:
\begin{align}
&[{\bm H}^{2\times 2}_{\rm eff}({\bm k},X_{m})]_{i,j} \equiv 
\sum_{{\bm q},{\bm q}^{\prime}} {\phi^{0}_{i}}^{*}({\bm q}) \!\ K({\bm q},{\bm q}^{\prime};{\bm k},X_{m}) \!\ 
\phi^{0}_{j}({\bm q}^{\prime}),   
\end{align}
($i,j=E_1,E_2$). Here $\phi^{0}_{j}({\bm q})$ is an eigenstate 
of the BS Hamiltonian at ${\bm k}=0$ and $X_{m}=0$:
\begin{align}
\sum_{ {\bm q}^{\prime} } K({\bm q},{\bm q}^{\prime};{\bm k}=0,X_{m}=0) \!\ \phi^{0}_{j}({\bm q}^{\prime}) = 
E_0 \!\ \phi^{0}_{j}({\bm q}) \nonumber 
\end{align}
with $j=E_{1},E_2$. From the Hellmann-Feynman theorem, the effective Hamiltonian at ${\bm k}=0$ 
with small $X_{m}$ is diagonal:
\begin{align}
{\bm H}^{2\times 2}_{\rm eff}({\bm k},X_{m}) = (E_0 + a_{+}X_{m}) {\bm \tau}_0 + a_- X_{m} {\bm \tau}_3 
+ {\cal O}(X^2_{m}), \label{22a} 
\end{align} 
with $2a_{\pm} = a_{E_1}\pm a_{E_2}$ and $a_{j} \equiv \langle \phi^{0}_j | (\partial_{X_{m}} K)_{|X_{m}=0} |\phi^0_j \rangle$.  
The ${\bm k}$-dependence of the Hamiltonian is constrained by the symmetries of $E_1$ and 
$E_2$ states (Table I), 
\begin{align}
&{{\bm H}^{2\times 2}_{\rm eff}}^{*}(-\sigma_{X}({\bm k}),X_{m}) =  
{{\bm H}^{2\times 2}_{\rm eff}}^{*}(-\sigma_{Y}({\bm k}),X_{m})  \nonumber \\ 
& = {\bm H}^{2\times 2}_{\rm eff}(C^{\pm}_4({\bm k}),X_{m})= {\bm \tau}_3 \!\  
{\bm H}^{2\times 2}_{\rm eff}({\bm k},X_{m})  \!\ {\bm \tau}_3,  
\end{align} 
and 
\begin{align}
&{{\bm H}^{2\times 2}_{\rm eff}}^{*}(-\sigma_{x(y)}({\bm k}),X_{m}) = {\bm H}^{2\times 2}_{\rm eff}(C^{2}_4({\bm k}),X_{m}) \nonumber \\ 
&={\bm H}^{2\times 2}_{\rm eff}({\bm k},X_{m}).    
\end{align}
These relations in combination with Eq.~(\ref{22a}) lead to 
\begin{align}
{\bm H}^{2\times 2}_{\rm eff}({\bm k},X_{m}) &= (E_0 + a_{+}X_{m} + a_0 k^2) 
{\bm \tau}_0 + a_- X_{m} {\bm \tau}_3 \nonumber \\
& \ \  + a_1 (k^2_x-k^2_y) {\bm \tau}_1 + a_2 k_x k_y {\bm \tau}_2  \nonumber \\
& \ \ + a_3 k^2 {\bm \tau}_3 + {\cal O}(X^2_{m},{\bm k}^3).  \label{23a} 
\end{align}
The same form also applies to the quadratic band touching formed by 
$A$ ($s$) and $B$ ($d$) states. One can have another symmetry-allowed form 
for exciton band touchings composed by other pairs. For $A$ and 
$E_{2}$ 
states (and also for $E_{1}$ and $B$ states), the band touching takes a form 
of 
\begin{align}
{\bm H}^{2\times 2}_{\rm eff}({\bm k},X_{m}) &=   (E_0 + b_{+}X_{m}) {\bm \tau}_0 + 
b_- X_{m} {\bm \tau}_3 \nonumber \\
& \ \ \  + b k_x {\bm \tau}_2 + b k_y {\bm \tau}_1 + {\cal O}(X^2_{m},{\bm k}^2).  \label{23b}  
\end{align} 
     
\subsection{M point}
The effective Hamiltonian for band touching at $M$ points can be derived in the same 
way. Consider $M=(\pi,0)$, whose group under which the BS Hamiltonian is invariant at this point 
is $m^{\prime}m^{\prime}2$  
(a subgroup of $4m^{\prime}m^{\prime}$): $m^{\prime}m^{\prime}2=\{E,C^2_4,\sigma_x T,\sigma_y T\}$. 
There are two distinct 1-dimensional IcREPs of $m^{\prime}m^{\prime}2$ (see Table.~II). 
They are $A$ and $B$,~\cite{s-bradley} in 
which eigenstates of the BS Hamiltonian at ${\bm k}=(\pi,0)$ is transformed with 
$s$ and $p_x$-wave symmetry respectively. 
From the character table, the ${\bm k}$-dependence of the effective 
Hamiltonian for this band touching is constrained as follows: 
\begin{align}
&{\bm H}^{2\times 2}_{\rm eff}(C_2({\bm k})) = {{\bm H}^{2\times 2}_{\rm eff}}^{*}(-\sigma_y({\bm k}))
={\bm \tau}_3 \!\ {\bm H}^{2\times 2}_{\rm eff}({\bm k})\!\ {\bm \tau}_3,  
\nonumber \\
&{{\bm H}^{2\times 2}_{\rm eff}}^{*}(-\sigma_x({\bm k})) = {\bm H}^{2\times 2}_{\rm eff}({\bm k}). \nonumber  
\end{align}
This leads to 
\begin{align}
{\bm H}^{2\times 2}_{\rm eff}({\bm k},X_{m}) &= (E_0 + c_{+}X_{m}) {\bm \tau}_0 
+ c_- X_{m} {\bm \tau}_3 \nonumber \\
& + c_1 k_x {\bm \tau}_2 + c_2 k_y {\bm \tau}_1 + {\cal O}(X^2_{m},{\bm k}^2) . \label{33a}
\end{align}

\begin{table}[h]
\caption{character table of $m'm'2$}
\begin{center}
\begin{ruledtabular}
\begin{tabular}{lcccc}
        & $E$ & $C_{4}^{2}$ & $\sigma_{x}T$ & $\sigma_{y}T$   \\
\colrule
   $A$ & $1$ &$1$ & $1$& $1$ \\ 
   $B$ & $1$ &$-1$ & $1$& $-1$ \\ 
\end{tabular}
\vspace{0.2cm}

\end{ruledtabular}
\end{center}

\end{table}

\section{The response function for the system with open boundary condition}

In the region with non-zero Chern integers in Fig.~3 in the main text, we calculate  
the response function for the system with open/periodic boundary condition in  
one ($x$)/the other ($y$) direction, to enumerate all possible collective 
excitations including edge and bulk modes. The formulation given below 
enables us to qualitatively discuss how gapless electronic 
edge mode affects the exciton edge modes. We begin with the interacting 
electron Hamiltonian with open boundary condition in $x$, ${\cal H}={\cal H}_0+{\cal V}$ 
with ${\cal H}_0 = \sum_{k_y} {\bm c}^{\dagger}_{k_y} {\bm H}_{sp}(k_y) {\bm c}_{k_y}$ and 
\begin{align} 
&{\cal V} = \frac{U}{2N} \sum_{k_{1,y},k_{2,y},q_y,\alpha,\beta} 
\sum^{N}_{x=1} \nonumber \\
& \ \ \ \ \ c^{\dagger}_{k_{1,y}+q_y,x,\alpha} c^{\dagger}_{k_{2,y}-q_y,x,\beta} 
c_{k_{2,y},x,\beta} c_{k_{1,y},x,\alpha}. \nonumber  
\end{align} 
Here ${\bm c}^{\dagger}_{k_y}$ takes a $2N$-component vector form,  
\begin{align}
&{\bm c}^{\dagger}_{k_y} \equiv \left(\begin{array}{ccccc} 
c^{\dagger}_{k_y,N,s\uparrow} & c^{\dagger}_{k_y,N,p\downarrow} & 
\cdots c^{\dagger}_{k_y,1,s\uparrow} & c^{\dagger}_{k_y,1,p\downarrow} 
\end{array}\right),   \nonumber 
\end{align}
where  
\begin{align}
&c^{\dagger}_{k_y,x,\alpha} = \frac{1}{\sqrt{N}} \sum^{N}_{y=1} e^{ik_y y} c^{\dagger}_{{\bm r},\alpha}, 
\end{align} 
with ${\bm r}\equiv (x,y)$, $x=1,2,\cdots,N$ and $\alpha=s\uparrow,p\downarrow$. ${\bm H}_{sp}(k_y)$ 
is a $2N\times2N$ Hermitian matrix, which is a Fourier transform of a tight-binding 
$sp$ model in the real space with open/periodic boundary in $x$/$y$ direction. 

The generalized random phase approximation~\cite{s-fetter,s-hanke} 
gives the response function 
$\chi^{R}({\bm r}_j,t_j;{\bm r}_m,t_m) \equiv -i\theta(t_j-t_m) \langle [{\cal O}_{H}({\bm r}_j,t_j),
{\cal O}_H({\bm r}_m,t_m)]\rangle$. The Fourier 
transform of the response function in the $y$ coordinate 
and time takes a $4N\times4N$ matrix form, 
\begin{widetext}
\begin{align}
\chi^{R}_{ab}(x_j,x_m;k_y,\omega) &\equiv \int d(t_j-t_m) \int d(y_j-y_m) 
e^{-ik_y (y_j-y_m) + i\omega (t_j-t_m)} \chi^{R}_{ab}({\bm r}_j,t_j;{\bm r}_m,t_m) 
= \chi^{T}_{ab}(x_j,x_m;k_y,i\omega_n=\omega+i\eta) \nonumber \\ 
{\bm \chi}^T_{ab}(k_y,i\omega_n) &= {\bm \Pi}^{U}_{ab}(k_y,i\omega_n) + U {\bm \Pi}^{U}_{a0}(k_y,i\omega_n) 
\Big[{\bm 1}_{N\times N} - U {\bm \Pi}^{U}_{00}(k_y,i\omega_n) 
\Big]^{-1}  {\bm \Pi}^{U}_{0b}(k_y,i\omega_n) \label{dyson} \\ 
{\bm \Pi}^{U}(k_y,i\omega_n) &\equiv {\bm \Pi}^{0}(k_y,i\omega_n)
 \Big[{\bm 1}_{4N\times 4N} + \frac{U}{2} {\bm \Pi}^{0}(k_y,i\omega_n) \Big]^{-1}, \label{bare} 
\end{align}
with $N\times N$ matrix ${\bm \chi}^T_{ab}(k_y,i\omega_n)$ as 
$[{\bm \chi}^T_{ab}(k_y,i\omega_n)]_{(x_j,x_m)} \equiv \chi^{T}_{ab}(x_j,x_m;k_y,i\omega_n)$ ($x_j,x_m=1,\cdots,N$). 
$N\times N$ matrices 
${\bm \Pi}^{U}_{ab}(k_y,i\omega_n)$ and ${\bm \Pi}^{0}_{ab}(k_y,i\omega_n)$ ($a,b=0,1,2,3$) comprise 
$4N\times 4N$ matrices ${\bm \Pi}^{U}(k_y,i\omega_n)$ and 
${\bm \Pi}^0(k_y,i\omega_n)$ respectively, 
\begin{align}
[{\bm \Pi}^{U}(k_y,i\omega_n)]_{(a,b)} &\equiv {\bm \Pi}^{U}_{ab}(k_y,i\omega_n), \nonumber \\
[{\bm \Pi}^{0}(k_y,i\omega_n)]_{(a,b)} &\equiv {\bm \Pi}^{0}_{ab}(k_y,i\omega_n). \nonumber 
\end{align}
A bare polarization function ${\bm \Pi}^{0}(k_y,i\omega_n)$ is given by single-particle 
electron eigenstates of ${\bm H}_{sp}(k_y)$:
\begin{align}
[{\bm \Pi}^{0}_{ab}(k_y,i\omega_n)]_{(x_j,x_m)} & \equiv \frac{1}{\beta} \frac{1}{N} 
\sum_{q_y} \sum_{i\epsilon_n} \sum_{t,s} \sum_{\alpha,\beta,\gamma,\delta} 
[{\bm \sigma}_a]_{\alpha\beta} [{\bm \sigma}_b]_{\gamma\delta}  
 \frac{\langle x_j,\beta | k_y+q_y,t\rangle 
\langle k_y+q_y,t | x_m,\gamma \rangle}{i(\epsilon_n+\omega_n)-{\cal E}_{k_y+q_y,t}+\mu} 
\frac{\langle x_m,\delta | q_y,s\rangle 
\langle q_y,s | x_j,\alpha \rangle}{i\epsilon_n-{\cal E}_{q_y,s}+\mu} \nonumber \\
& = \frac{1}{N} 
\sum_{q_y} \sum_{t,s} \sum_{\alpha,\beta,\gamma,\delta} \frac{n_{F}({\cal E}_{k_y+q_y,t}) - 
n_F({\cal E}_{q_y,s})}{-i\omega_n+({\cal E}_{k_y+q_y,t}-{\cal E}_{q_y,s})} 
\!\ [{\bm \sigma}_a]_{\alpha\beta} [{\bm \sigma}_b]_{\gamma\delta} \nonumber \\  
& \hspace{2cm} \times \langle x_j,\beta | k_y+q_y,t\rangle 
\langle k_y+q_y,t | x_m,\gamma \rangle 
\langle x_m,\delta | q_y,s\rangle 
\langle q_y,s | x_j,\alpha \rangle  \label{T=0}  
\end{align}  
$n_F({\cal E})$ denotes the Fermi distribution function. In the following, we assume $T=0$. 
$|k_y,t\rangle$ and ${\cal E}_{k_y,t}$ 
are the single-particle eigenstate and eigenenergy of ${\bm H}_{sp}(k_y)$. This includes 
both bulk (`$t=b$') and edge states (`$t=e$').  

We decompose the bare polarization function into three parts, depending on whether 
the polarization is induced by bulk states or edge states: 
\begin{align}
{\bm \Pi}^{0}(k_y,i\omega_n) = {\bm \Pi}^{0,(b,b)}(k_y,i\omega_n) + 
 {\bm \Pi}^{0,(b,e)}(k_y,i\omega_n) + {\bm \Pi}^{0,(e,e)}(k_y,i\omega_n). 
 \end{align} 
\end{widetext}
Namely, in ${\bm \Pi}^{0,(b,b)}$, the summations over 
the single-particle states in Eq.~(\ref{T=0}) (the summations over `$s$' and `$t$' in Eq.~(\ref{T=0})) are over 
the bulk states. In  ${\bm \Pi}^{0,(e,e)}$, the summations are 
only over the edge states. In ${\bm \Pi}^{0,(b,e)}$, the summation over 
the particle states (`$t$' in Eq.~(\ref{T=0})) is over the bulk (edge) states, if that over 
the hole states (`$s$' in Eq.~(\ref{T=0})) is taken over the edge (bulk) states. 
For given $k_y$, the number of the bulk states is much larger 
than that of the edge states. Therefore, ${\bm \Pi}^{0,(b,b)}$ has a major contribution to 
${\bm \Pi}^{0}$. ${\bm \Pi}^{0,(b,e)}$ and ${\bm \Pi}^{0,(e,e)}$ 
have secondary roles, which we will discuss later.     

For $\omega$ below the electron-hole continuum of the 
bulk states, ${\bm \Pi}^{0,(b,b)}(k_y,\omega)$ is a Hermitian matrix.  
${\bm \Pi}^{U}$ (or ${\bm \chi}^{R}$) made only out of such 
${\bm \Pi}^{0,(b,b)}$ by Eqs.~(\ref{bare},\ref{dyson}) can be 
diagonalized by a unitary matrix, unless one of its eigenvalues has a pole. Namely,
\begin{eqnarray}
\left[\bm{\Pi}^{U}\left(k_{y},\omega\right)\right]^{-1} |v_{a}(k_y,\omega) \rangle 
= |v_{a}(k_y,\omega) \rangle {\cal L}^D_{a} (k_y,\omega)  
\end{eqnarray}
where $|v_{a}(k_y,\omega)\rangle$ with $a=1,\cdots,4N$ form an orthonormal basis for 
those $\omega$ with ${\cal L}^D_{a}(k_y,\omega) \ne 0$ for any $a$. 
Like in the bulk calculation (see main text), 
each eigenvalue has at most one pole below the electron-hole continuum  
of the bulk states. The pole is determined by the zero of the 
eigenvalue: ${\cal L}^D_{a}(k_y,E_{a,k_y})=0$. 
$E_{a,k_y}$ gives an energy-momentum dispersion of exciton 
states with a given momentum $k_y$. Such exciton states 
inlcude both bulk exciton states and edge exciton states. 

\begin{figure}[t]
	\includegraphics[width=0.9\columnwidth]{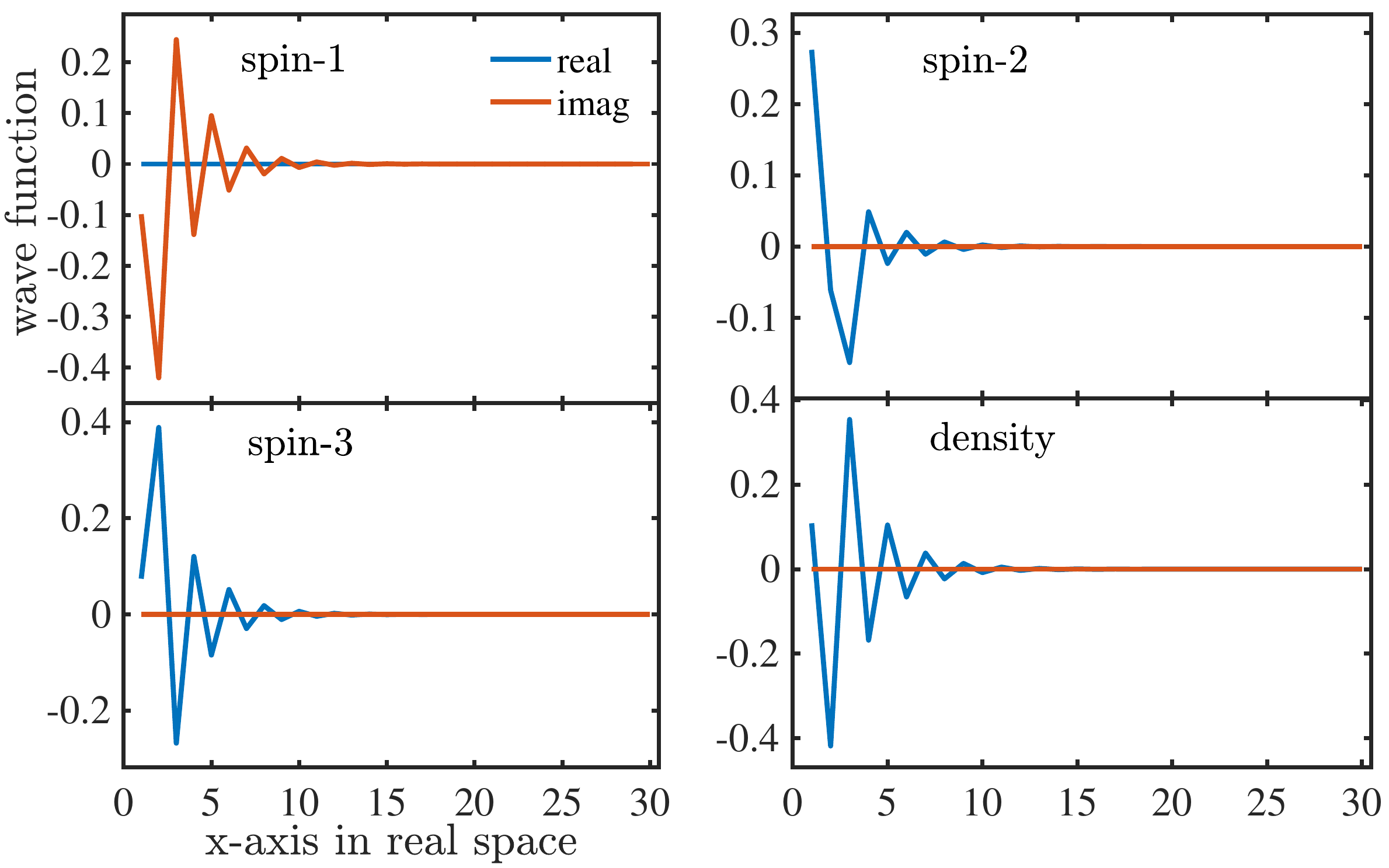}
  	\includegraphics[width=0.9\columnwidth]{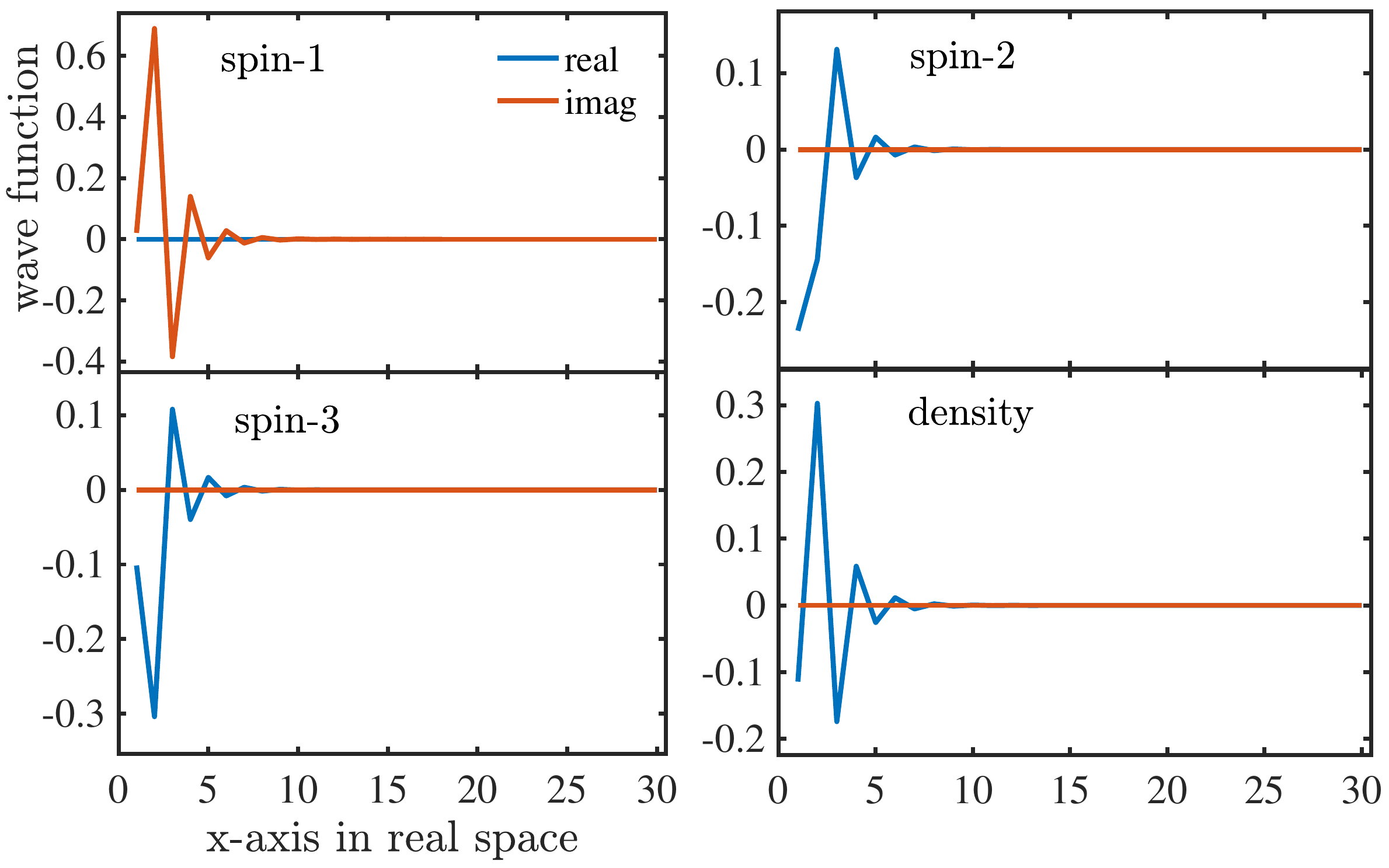}
	\caption{(color online) Real (blue) and imaginary (red) parts in spin 1 (${\bm \sigma}_1$), 2 (${\bm \sigma}_2$), 
3 (${\bm \sigma}_3$) and density (${\bm \sigma}_0$) components of eigen wavefunctions ($|\tilde{v}_{a}(k_y)\rangle 
\equiv |v_{a}(k_y,E_{a,k_y})\rangle$) which correspond to those in-gap exciton states at $k_y=0.13\pi$ 
in Fig.~1 in the main text (ingap states with blue color chiral dispersion). They are plotted as a function of 
the spatial coordinate $x$. $x=0$ and $x=30$ correspond to the left-hand-side and right-hand-side spatial boundary. 
The upper four are for Fig.~1(a) and the lower four are for Fig.~1(b) in the main text.}
	\label{Fig:wave}
\end{figure}

Firstly, we confirm that exciton states thus obtained completely reproduce the 
lowest two exciton bulk bands obtained from the calculation with periodic boundary 
conditions both in $x$ and $y$ direction (main text). 
Besides, we observe that, when an indirect band 
gap opens between the lowest two exciton bulk bands, new exciton 
states appear inside the gap. Eigenvalues ($E_{a,k_y}$) for these in-gap exciton 
states form chiral energy-momentum 
dispersions as a function of $k_y$, connecting the two exciton bulk bands 
(Fig.~1 in the main text). 
The number of the chiral dispersions inside the gap including its sign 
turns out to be identical to the Chern integer of the lowest exciton bulk band. 
The eigen wavefunctions $(|\tilde{v}_a(k_y)\rangle \equiv 
|v_{a}(k_y,\omega=E_{a,k_{y}})\rangle)$ which correspond to these in-gap exciton states, 
are spatially localized at the boundaries, having weight in the pseudospin component of 
${\bm \sigma}_1$ and ${\bm \sigma}_2$ (Fig.~\ref{Fig:wave}). These observations 
justify the presence of 
the chiral exciton edge modes of the topological origin in $sp$ model with 
short-range Coulomb interaction, which interact with light. 

Let us next discuss effects of ${\bm \Pi}^{0,(b,e)}$ and ${\bm \Pi}^{0,(e,e)}$. The 
electronic edge mode is gapless, so that these two have both Hermitian and anti-Hermitian parts. 
Nonetheless, any matrix element of the anti-Hermitian part of 
${\bm \Pi}^{0,(b,e)}$ is negligibly small in the thermodynamic limit. 
Namely, the bulk states are extended in space, so that 
an integrand in Eq.~(\ref{T=0}) for ${\bm \Pi}^{0,(b,e)}$ is 
at most on the order of $1/N$, i.e.,  
\begin{align}
&\langle x_j,\beta | k_y+q_y,t=b\rangle 
\langle k_y+q_y,t=b | x_m,\gamma \rangle \nonumber \\ 
&\ \ \times\langle x_m,\delta | q_y,s=e\rangle 
\langle q_y,s=e | x_j,\alpha \rangle < {\cal O} \Big(\frac{1}{N}\Big). 
\end{align} 
(`$t=b$' and `$s=e$' means that the single-particle state for $t$ is from the bulk states 
and that for $s$ is from the edge state). Meanwhile, matrix elements 
of an anti-Hermitian part of 
${\bm \Pi}^{0,(e,e)}$, which are also spatially localized at the boundaries, 
are on the order of $1$.

The effect of ${\bm \Pi}^{0,(e,e)}(k_y,\omega)$ is two-folded, 
depending on an energy region $\omega$. 
In low-energy region, the Hermitian part of 
${\bm \Pi}^{0,(e,e)}$ leads to the one-dimensional chiral phason mode. Being a 
well-defined low-energy collective bosonic excitation, the phason mode has 
a level crossing (and consequently level repulsion) with low-energy chiral exciton edge 
modes (Fig.~4 in the main text). The 
level repulsion changes total number of low-energy chiral dispersions of edge 
excitons which go across the indirect gap between the lowest two exciton bulk bands. 
For $0<m<2t$ $(-2t<m<0)$, total number of the chiral dispersions becomes 
${\mathbb Z}-1$ $({\mathbb Z}+1)$ instead of ${\mathbb Z}$, where ${\mathbb Z}$ is the 
Chern integer of the lowest bulk exciton band. 

In high-energy region, the anti-Hermitian part of ${\bm \Pi}^{0,(e,e)}$ results in a finite 
region of an edge electron-hole continuum. The electronic edge mode in the high energy 
region generally has a curvature in its energy-momentum dispersion. 
Electron-hole excitations along such electronic edge mode 
form a finite region of a new electron-hole continuum (below the 
electron-hole continuum of the bulk states). When the chiral exciton edge modes share 
energy and momentum with the edge electron-hole continuum, they acquire 
a finite life time (Fig.~4 in the main text).  


\bibliographystyle{apsrev4-1}
\bibliography{refs}


\end{document}